\documentclass[conference]{IEEEtran}
\usepackage{amsmath}
\usepackage{amssymb}
\usepackage[dvips]{graphicx}
\usepackage{epsfig}

\usepackage{amsmath}
\usepackage{amssymb}
\usepackage[dvips]{graphicx}
\usepackage{setspace}
\usepackage{epsfig}

\newcommand{\Ei}{\textrm{Ei}}

\newcommand{\E}{\mathbb{E}}

\newcommand{\gwbar}{\overline{\gamma}_E}
\newcommand{\gmbar}{\overline{\gamma}_M}
\newcommand{\gpbar}{\overline{\gamma}_P}

\newcommand{\ud}{{\mathrm{d}}}

\newtheorem{rem}{Remark}

\begin{document}


\title{Optimal Power Allocation for Secrecy Fading Channels Under Spectrum-Sharing Constraints}



%
\author{\authorblockN{Junwei Zhang and Mustafa Cenk Gursoy}
\authorblockA{Department of Electrical Engineering\\
University of Nebraska-Lincoln, Lincoln, NE 68588\\ Email:
junwei.zhang@huskers.unl.edu, gursoy@engr.unl.edu}}

\maketitle
\begin{abstract}\footnote{This work was supported by the National Science Foundation under Grant CCF -- 0546384 (CAREER).}
In the spectrum-sharing technology, a secondary user may utilize the
primary user's licensed band as long as its interference to the
primary user is below a tolerable value. In this paper, we consider a scenario in which a secondary user is operating in the presence of both a primary user and an eavesdropper. Hence, the secondary user has both interference limitations and security considerations. In such a scenario, we study the
secrecy capacity limits of opportunistic spectrum-sharing channels
in fading environments and investigate the optimal power allocation
for the secondary user under average and peak received power constraints
at the primary user with global channel side information (CSI). Also, in the absence of
the eavesdropper's CSI, we study optimal power allocation under an average
power constraint and propose a suboptimal on/off power control
method.

\emph{Index Terms:} Spectrum sharing, cognitive radio,
physical-layer security, fading channel, power control.
\end{abstract}

\section{introduction}

The proliferation of wireless systems and services has increased the need to efficiently use the scarce spectrum in wireless applications. This, together with the recently observed fact that the spectrum resources are not being utilized effectively, have spurred much interest in the study of cognitive radio networks. In 
cognitive radio networks, the secondary users may be allowed to transmit
concurrently in the same frequency band with the primary users as
long as the resulting interference power at the primary receivers is
kept below the interference temperature limit \cite{Haykin}.  A
significant amount of work has been done to study the transmitter
design under such interference constraints, e.g., in \cite{Ghasemi}
and \cite{Musavian} for the fading channel, in \cite{Rzhang} for the
multiple-input multiple-output (MIMO) channel, in  \cite{Mietzner}
for the relay channel.

On the other hand, the broadcast nature of wireless transmissions
allows for the signals to be received by all users within the
communication range, making wireless communications vulnerable to
eavesdropping. The problem of secure transmission in the presence of
an eavesdropper was first studied from an information-theoretic
perspective in \cite{wyner} where Wyner considered a wiretap channel
model. Wyner showed that secure communication is possible without
sharing a secret key if the eavesdropper's channel is a degraded
version of the main channel, and identified the rate-equivocation
region and established the secrecy capacity of the degraded discrete
memoryless wiretap channel. The secrecy capacity is defined as the
maximum achievable rate from the transmitter to the legitimate
receiver, which can be attained  while keeping the eavesdropper
completely ignorant of the transmitted messages. Later, Wyner's
result was extended to the Gaussian channel in \cite{cheong} and
recently to fading channels in \cite{Liang} and \cite{Gopala}. In
addition to the single antenna case, secrecy in multi-antenna models
was addressed in \cite{shafiee} and \cite{khisti}.   Cooperative
relaying under secrecy constraints was also recently studied in
\cite{dong}--\cite{jzhang1}.

In this paper, we consider a scenario in which second users communicate in the presence of a primary user and an eavesdropper. Hence, secondary users need to both control the interference levels on the primary user and send the information securely. Hence, we combine the challenges seen in studies of cognitive radio networks and
information-theoretic security.  We note that, despite its practical relevance, security considerations in cognitive transmissions have received relatively little attention in
research. Capacity of cognitive interference channel with secrecy is
studied in \cite{Liang1}. Recently, \cite{Pei} has studied
secure communication over MISO cognitive radio channels. The secrecy
capacity of the channel is characterized, and finding the
capacity-achieving transmit covariance matrix under the joint
transmit power and interference power constraints is formulated as a
quasiconvex optimization problem.

Our contributions in this paper are as follows. We initially assume that the transmitter has global channel side information (CSI), i.e., perfectly knows the fading coefficients of all channels, and we study the secrecy capacity limits of
opportunistic spectrum-sharing channels in fading environments and
identify the optimal power allocation for the secondary user under
average and peak received power constraints at the primary user.
Subsequently, we consider the case in which the eavesdropper's CSI is unavailable at the source. In this scenario, we
study the optimal power allocation under average power constraints,
and propose a simplified on/off power control method.

\section{Channel Model}
\begin{figure}
\includegraphics[width=0.5\textwidth]{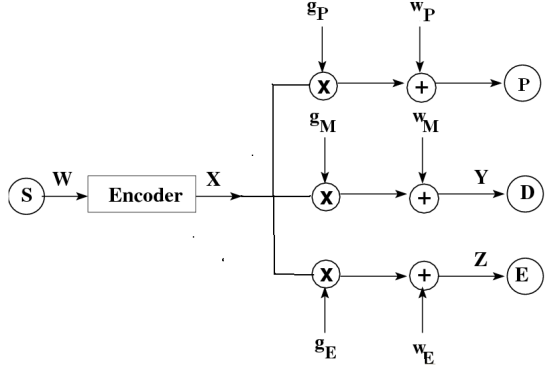} \hspace{0.5in}
\caption{Channel Model\label{model}}
\end{figure}
As depicted in
Fig.\ref{model}, we consider a cognitive radio channel model
with a secondary user source $S$, a primary user $P$, a secondary
user destination $D$,  and an eavesdropper $E$. In this model, the source $S$ tries to transmit
confidential messages to destination $D$ on the same band as the primary user's
while keeping the interference on the primary user below some predefined
interference temperature limit and keeping the eavesdropper $E$ ignorant of the information.
 During any coherence interval $i$,
the signal received by the destination and the eavesdropper are
given, respectively, by
\begin{eqnarray}
y(i)&=&g_{M}(i)x(i)+w_{M}(i),\\
z(i)&=&g_{E}(i)x(i)+w_{E}(i),
\end{eqnarray}
where $g_{M}(i),g_{E}(i)$ are the channel gains from the secondary
source to the secondary receiver (main channel) and from the secondary source to the eavesdropper
(eavesdropper channel), respectively, and $w_{M}(i),w_{E}(i)$
represent the i.i.d additive Gaussian noise with zero-mean and unit-variance at
the destination and the eavesdropper, respectively. We denote the
fading power gains of the main and eavesdropper channels by $h_M(i)
= |g_M(i)|^2$ and $h_E(i) = |g_E(i)|^2$, respectively. Similarly, we
denote the channel gain from the secondary source to the primary
receiver by $ g_{P}(i)$ and  its fading power gain by $h_P(i) =
|g_P(i)|^2$. We assume that both channels experience block fading, i.e.,
the channel gains remain constant during each coherence
interval and change independently from one coherence interval to the
next. The fading process is assumed to be ergodic with a bounded
continuous distribution. Moreover, the fading coefficients of the
destination and the eavesdropper in any coherence interval are
assumed to be independent of each other.

Since transmissions pertaining to the secondary user should not
harm the signal quality at the receiver of the primary user, we
impose constraints on the received-power at the primary user $P$.
Hence, denoting the average and peak received-power values by
$Q_{avg}$ and $Q_{peak}$, respectively, we define the corresponding
constraints as:
\begin{align}
\E_{h_M,h_E,h_P}\{P(h_M,h_E,h_P)h_P\}\leq Q_{avg}\\
\intertext{and}
P(h_M,h_E,h_P)h_P\leq Q_{peak}, ~~~\forall h_M,h_E,h_P.
\end{align}
Note that $Q_{avg}$ can be seen as a long-term average received power constraint. Additionally, although we call $Q_{peak}$ as the peak received-power constraint, it is actually a peak constraint on the average instantaneous received power and can be regarded as a short-term constraint.


\section{Power Allocation under Average Received-Power Constraints}
In a fading environment, following the the same line of development
as in \cite{Gopala}, it is straightforward but tedious to show that
the channel capacity is achieved by optimally distributing the
transmitted power over time such that the  primary user received
power constraint is met. By assuming that $h_M$, $h_E$, and $h_P$ are
independent of each other and global CSI is available, the secrecy
capacity under an average received power constraint is the solution to the following optimization problem,
\begin{align}
&\max_{P(h_M,h_E,h_P)\geq 0}\int\int\int\Big[ \log \left( 1 + h_M
P(h_M,h_E,h_P) \right)\nonumber \\&\hspace{3.5cm}-\log \left( 1 + h_E
P(h_M,h_E,h_P) \right)
\Big]^+\nonumber\\
&\hspace{3.3cm}\times f(h_M)f(h_E)f(h_P)\ud h_M \ud h_E\ud h_P \nonumber \\
&s.t~~~\int\int\int h_PP(h_M,h_E,h_P)\nonumber \\
&~~~~~~~~~~~~~~\times f(h_M)f(h_E)f(h_P)\ud h_M \ud h_E\ud h_P\leq
Q_{avg}\label{powercons}
\end{align}
where $[x]^{+} = \max\{0,x\}$. To find the optimal power allocation
$P(h_M,h_E,h_P)$, we form the Lagrangian:
\begin{align}
&L(P,\lambda)=\int\int\int\Big[ \log \left( 1 + h_M P(h_M,h_E,h_P)
\right)\nonumber
\\
&-\log \left( 1 + h_E P(h_M,h_E,h_P)\right)
\Big]^+f(h_M)f(h_E)f(h_P)\ud h_M \ud h_E\ud h_P \nonumber \\
&-\lambda \Big(\int\int\int h_PP(h_M,h_E,h_P)\nonumber\\
&~~~~\times f(h_M)f(h_E)f(h_P)\ud h_M \ud h_E\ud h_P- Q_{avg}\Big).
\end{align}
By using the Lagrangian maximization approach, we get the
following optimality condition:
\begin{align}\label{Optcond}
&\frac{\partial L(P,\lambda)}{\partial P(h_M,h_E,h_P)}\nonumber \\
&=(\frac{h_M}{1+h_M P(h_M,h_E,h_P)} - \frac{h_E}{1+h_E
P(h_M,h_E,h_P)} - \lambda h_P) \nonumber\\
&\hspace{.5cm}\times f(h_M)f(h_E)f(h_P)~=~0.
\end{align}
Solving (\ref{Optcond}) with the constraint  $P(h_M,h_E,h_P)\geq 0$
yields the optimal power allocation policy at the transmitter as
\begin{align}\label{Lar}
P(h_M,h_E,h_P) = \frac{1}{2} \Bigg[ \sqrt{ \left( \frac{1}{h_E} -
\frac{1}{h_M} \right)^2 +  \frac{4}{\lambda h_P} \left(
\frac{1}{h_E} -
 \frac{1}{h_M} \right)}\nonumber \\ - \left( \frac{1}{h_M} + \frac{1}{h_E} \right)
 \Bigg]^+,
\end{align}
where $\lambda$ is a constant that is introduced to satisfy the receive power
constraint (\ref{powercons}) at the primary user.

\begin{rem}
It is easy to see that when $h_E>h_M$, $P(h_M,h_E,h_P)=0$, which is
in accordance with our intuition. Transmitter only spends power for
transmission when the main channel is better than the eavesdropper's
channel. With little calculation, we can also see that when $h_P
> \frac{h_M-h_E}{\lambda}$, we have $P(h_M,h_E,h_P)=0$. Thus, the
power allocation can be rewritten as
\begin{align}\label{ave}
P(h_M,h_E,h_P)= \left\{
\begin{array}{ll}
\frac{1}{2} \Bigg[ \sqrt{ \left( \frac{1}{h_E} - \frac{1}{h_M}
\right)^2 +  \frac{4}{\lambda h_P} \left( \frac{1}{h_E} -
 \frac{1}{h_M} \right)} \\ - \left( \frac{1}{h_M} + \frac{1}{h_E} \right)
 \Bigg] ~~~~~~~ \frac{h_M-h_E}{h_P}>\lambda
\\
0~~~~~~~~~~~~~~~~~~~~~~~~~~  \frac{h_M-h_E}{h_P} \leq\lambda
\end{array}\right..
\end{align}
\end{rem}

\begin{rem}
From the expression of the optimal power allocation obtained in
(\ref{Lar}), we can easily see that more transmission power is used
when either $h_M$ increases or $h_P$ decreases. Also the derivative of
(\ref{Lar}) with regard to $h_E$ is
\begin{align}
-\frac{1}{2h_E^2}\Big[\frac{\frac{1}{h_E}-\frac{1}{h_M}+\frac{2}{\lambda
h_P}}{\sqrt{ \left( \frac{1}{h_E} - \frac{1}{h_M} \right)^2 +
\frac{4}{\lambda h_P} \left( \frac{1}{h_E} -
 \frac{1}{h_M} \right)}}-1\Big].
\end{align}
 We can see that the derivative is negative, so $P(h_M,h_E,h_P)$
decreases when $h_E$ increases. These observations are also intuitively appealing. The secondary user takes advantage of the weak link between its
transmitter and the primary receiver, and the stronger main channel.
Also, a weaker eavesdropper's channel is preferred for secure
message transmission.
\end{rem}

\begin{rem}
When there is no eavesdropper, the channel is the standard cognitive
radio channel. By letting $h_E=0$ in (\ref{Optcond}) and solving the
problem, we can obtain the optimal power allocation as $(\frac{1}{\lambda
h_P}-\frac{1}{h_M})^+$, which has also been shown in
\cite{Ghasemi} and \cite{Musavian}.
\end{rem}

\begin{rem}
When there is no primary user, the channel is the standard secrecy
fading channel. By
replacing $h_P$ with $1$ in (\ref{powercons}) and correspondingly
replacing $h_P$ with $1$ in (\ref{Lar}), we get the optimal power
allocation for the fading secrecy channel given in
\cite{Gopala}.
\end{rem}

\section{Power Allocation under both Average and Peak Received-Power Constraints}
The average received power constraint is reasonable when the primary
user's QoS is determined by the average long-term interference. However, we note that
in many cases, the primary user's QoS is also limited by the
instantaneous interference at the primary receiver. With this motivation, we in this section
study the power allocation under both average and peak received power
constraints.

We first introduce  a real-valued function $\beta$ which is defined
as
\begin{align} \label{beta}
\beta ^2 \triangleq \frac{Q_{peak}}{h_P}-P(h_M,h_E,h_P).
\end{align}
To satisfy the peak power constraint, the right-hand side of
(\ref{beta}) must be nonnegative over all the possible values of the channel gain.
Using (\ref{beta}), we form an equivalent problem of
(\ref{powercons}), which contains an equality constraint for the
peak power.

\begin{align}
&\max_{P(h_M,h_E,h_P)\geq 0,\beta}\int\int\int\Big[ \log \left( 1 +
h_M P(h_M,h_E,h_P) \right)\nonumber \\&\hspace{3.5cm}-\log \left( 1 +
h_E P(h_M,h_E,h_P) \right)
\Big]^+\nonumber\\
&\hspace{3.3cm}\times f(h_M)f(h_E)f(h_P)\ud h_M \ud h_E\ud h_P\\
&s.t~~~\int\int\int h_PP(h_M,h_E,h_P)\nonumber \\
&~~~~~~~~~~~~~~\times f(h_M)f(h_E)f(h_P)\ud h_M \ud h_E\ud h_P\leq
Q_{avg}\ \\
&\text{and}~~~~~~~~~~~\beta^2+P(h_M,h_E,h_P)=\frac{Q_{peak}}{h_P}.\label{powerconsp}
\end{align}

Now, the Lagrangian becomes
\begin{align}
&L(P,\lambda)=\int\int\int\Big[ \log \left( 1 + h_M P(h_M,h_E,h_P)
\right)\nonumber
\\
&-\log \left( 1 + h_E P(h_M,h_E,h_P)\right)
\Big]^+f(h_M)f(h_E)f(h_P)\ud h_M \ud h_E\ud h_P \nonumber \\
&-\lambda \Big(\int\int\int h_PP(h_M,h_E,h_P)\nonumber\\
&~~~~\times f(h_M)f(h_E)f(h_P)\ud h_M \ud h_E\ud h_P-
Q_{avg}\Big)\nonumber\\
&-\lambda_0\Big( \beta^2+P(h_M,h_E,h_P)-\frac{Q_{peak}}{h_P}\Big).
\end{align}
Setting each of the partial derivatives of the Lagrangian with respect to $P$ and
$\beta$ to zero, we obtain, respectively, the necessary conditions
for the optimal solution to  problem (\ref{powerconsp}) as
\begin{align}
\frac{h_M}{1+h_M P(h_M,h_E,h_P)} - &\frac{h_E}{1+h_E P(h_M,h_E,h_P)}
- \lambda h_P-\lambda_0=0\label{cond1} \\
&2\beta\lambda_0=0.\label{cond2}
\end{align}

Note that (\ref{cond2}) implies either $\beta=0$ or $\lambda_0=0$.
$\beta=0$ means that the peak power constraint is active and hence,
the optimal transmission power in this case is given by (\ref{pp})
\begin{align}\label{pp}
P(h_M,h_E,h_P)=\frac{Q_{peak}}{h_P}.
\end{align}
On the other hand, $\lambda_0=0$ in (\ref{cond2}) means that the
peak transmission power constraint is inactive and it can be
ignored. Solving (\ref{cond1}) with $\lambda_0=0$, we get the
expression for the optimal transmitter power as  $$ \frac{1}{2}
\Bigg[ \sqrt{ \left( \frac{1}{h_E} - \frac{1}{h_M} \right)^2 +
\frac{4}{\lambda h_P} \left( \frac{1}{h_E} -
 \frac{1}{h_M} \right)} - \left( \frac{1}{h_M} + \frac{1}{h_E} \right)
 \Bigg]^+,$$ which is the same expression as in (\ref{Lar}) obtained when there is only an average received power constraint.
Combining the two cases, the optimal power allocation under
both average and peak power constraints becomes
\begin{align}\label{ppa}
&\hspace{1cm}P(h_M,h_E,h_P)=\min \Bigg(\frac{Q_{peak}}{h_P}, \nonumber \\
&\hspace{-1cm}\frac{1}{2} \Bigg[ \sqrt{ \left( \frac{1}{h_E} - \frac{1}{h_M}
\right)^2 +  \frac{4}{\lambda h_P} \left( \frac{1}{h_E} -
 \frac{1}{h_M} \right)} - \left( \frac{1}{h_M} + \frac{1}{h_E} \right)
 \Bigg]^+ \Bigg)
\end{align}
where  $\lambda$ is a constant with which the average power
constraint is satisfied. We should note that $\lambda$ here is
generally not the same as $\lambda$ in the optimal power allocation in
(\ref{Lar}).

\begin{rem}
We can see from (\ref{ppa}) with little computation that when the
condition
\begin{align}
\frac{1}{\frac{h_E}{h_P}+1/Q_{peak}}-\frac{1}{\frac{h_M}{h_P}+1/Q_{peak}}>\lambda
Q_{peak}^2
\end{align}
is satisfied, we have $P(h_M,h_E,h_P)=\frac{Q_{peak}}{h_P}$
\end{rem}

\section{Power Allocation without Eavesdropper's CSI}

Since eavesdropping is a passive operation (i.e., does not involve any transmission), the source may not be able to get
the CSI of the eavesdropper's channel in certain circumstances. With this motivation, we in this section study the
optimal power allocation when the source knows only $h_M$ and
$h_P$. To simplify the analysis, we consider only average receive
power constraints here.

\subsection{Optimal Power Allocation}
 Based on the results of \cite{Gopala}, the
secrecy capacity in this case is the solution of the following
optimization problem:
\begin{align}
&\max_{P(h_M,h_P)\geq 0}\int\int\int\Big[ \log \left( 1 + h_M
P(h_M,h_P) \right)\nonumber \\&\hspace{3.5cm}-\log \left( 1 + h_E
P(h_M,h_P) \right)
\Big]^+\nonumber\\
&\hspace{3.3cm}\times f(h_M)f(h_E)f(h_P)\ud h_M \ud h_E\ud h_P \nonumber \\
&s.t~~~\int\int\int h_PP(h_M,h_P)\nonumber \\
&~~~~~~~~~~~~~~\times f(h_M)f(h_E)f(h_P)\ud h_M \ud h_E\ud h_P\leq
Q_{avg}.\label{powercons1}
\end{align}
Similarly, using the Lagrangian approach, we get the optimal
condition as
\begin{align}\label{optm}
&\frac{h_M \Pr \left(h_E \leq h_M \right)}{1+h_M P(h_M,h_P)}
\nonumber
\\- \int_{0}^{h_M}& \left( \frac{h_E}{1+h_E P(h_M,h_P)} \right) f(h_E)\ud
h_E - \lambda h_P~=~ 0,
\end{align}
where $\lambda$ is a constant that satisfies the power constraints
in (\ref{powercons1}) with equality. By solving (\ref{optm}), we can
get the optimal transmit power allocation $P(h_M,h_P)$. If the
obtained  value turns out to be negative, then the optimal value of
$P(h_M,h_P)$ is equal to 0. The exact solution to this optimization
problem depends on the fading distributions.

If Rayleigh fading scenario is considered with ${\mathbb E}\{h_M\}=
\gmbar$, ${\mathbb E}\{ h_E\}=\gwbar$  and ${\mathbb E}\{
h_P\}=\gpbar$ , then the optimal power allocation is the solution of
the following equation:
\begin{align}\label{rayleighoptm}
&\left( 1 - e^{-(h_M/\gwbar)} \right) \left( \frac{h_M}{1+ h_M
P(h_M,h_P)} \right)\lambda h_P \nonumber\\
& - \frac{\left( 1 - e^{-(h_M/\gwbar)} \right)}{P(h_M,h_P)} +
\frac{\exp \left( \frac{1}{\gwbar P(h_M,h_P)} \right)} {\gwbar
(P(h_M,h_P))^2} \Bigg[ \Ei \left(\frac{1}{\gwbar P(h_M,h_P)}\right)
\nonumber\\&- \Ei \left(\frac{h_M}{\gwbar} + \frac{1}{\gwbar
P(h_M,h_P)} \right) \Bigg]
 ~=~ 0
\end{align}
where $ \Ei (x) ~=~ \int_{x}^{\infty} \frac{e^{-t}}{t} ~\ud t$ is
the exponential integral function.  Again, if there is no positive
solution to (\ref{rayleighoptm}), the optimal $P(h_M,h_P)=0$.

\subsection{On/Off power control}
As seen above, the computation of the optimal power allocation
is in general complicated. In this section, we use  a simplified
suboptimal on/off power control method \cite{Gopala}. That is,  the
source sends information only when the channel gain $h_M$ exceeds a
pre-determined constant threshold $\tau > 0$. Moreover, when
$h_M>\tau$, the transmitter always uses the same power level $P$ .
It is easy to compute  that the constant power level used for
transmission should be
\begin{align}
P=\frac{Q_{avg}}{\gpbar \Pr (h_M > \tau)}.
\end{align}
For the Rayleigh fading scenario for which $f(h_M)=\frac{1}{\gmbar}
e^{-(h_M/\gmbar)}$, we get
 \begin{align}
 P=\frac{Q_{avg}}{\gpbar }e^{(\tau/\gmbar)}.
\end{align}
Then, the secrecy rate can be computed as
\begin{align}
&R_s= \int_{0}^{\infty} \int_{\tau}^{\infty} \left[ \log \left( 1 +
h_M P \right) -  \log \left( 1 + h_E P \right)
\right]^{+}\nonumber\\
&\hspace{2cm}\times f(h_M)
f(h_E) \ud h_M \ud h_E \nonumber\\
&=e^{-(\tau/\gmbar)} \log \left( 1 + \tau \frac{Q_{avg}}{\gpbar}
e^{(\tau/\gmbar)} \right) + \exp \left(\frac{1}{\gmbar
\frac{Q_{avg}}{\gpbar}
e^{(\tau/\gmbar)}}\right)\nonumber\\
&\Ei \left( \frac{\tau}{\gmbar} +\frac{1}{\gmbar
\frac{Q_{avg}}{\gpbar} e^{(\tau/\gmbar)}} \right)  + \exp \left(
\frac{1}{\gwbar \frac{Q_{avg}}{\gpbar} e^{(\tau/\gmbar)}} -
\frac{\tau}{\gmbar}
\right)\nonumber\\
& \left[ \Ei \left(\frac{\tau}{\gwbar} + \frac{1}{\gwbar
\frac{Q_{avg}}{\gpbar} e^{(\tau/\gmbar)}} \right) - \Ei
\left(\frac{1}{\gwbar \frac{Q_{avg}}{\gpbar} e^{(\tau/\gmbar)}}
\right)
\right]\nonumber\\
 &- \exp \left( \frac{\left[ \frac{1}{\gmbar} +
\frac{1}{\gwbar} \right]}{\frac{Q_{avg}}{\gpbar}
e^{(\tau/\gmbar)}}\right) \Ei \left( \left[\frac{1}{\gmbar} +
\frac{1}{\gwbar} \right] \left[\tau +
\frac{1}{\frac{Q_{avg}}{\gpbar} e^{(\tau/\gmbar)}} \right] \right).
\end{align}
Note that the secrecy rate depends on  the threshold $\tau$. Hence,
we can get the maximum achievable secrecy rate under the on/off power
control policy by optimizing the threshold $\tau$.

\section{Numerical Results}

In this section, we numerically illustrate the secrecy rate studied
in this paper. In all simulations, we assume that the fading is Rayleigh distributed.

\begin{figure}
\begin{center}
\includegraphics[width = 0.45\textwidth]{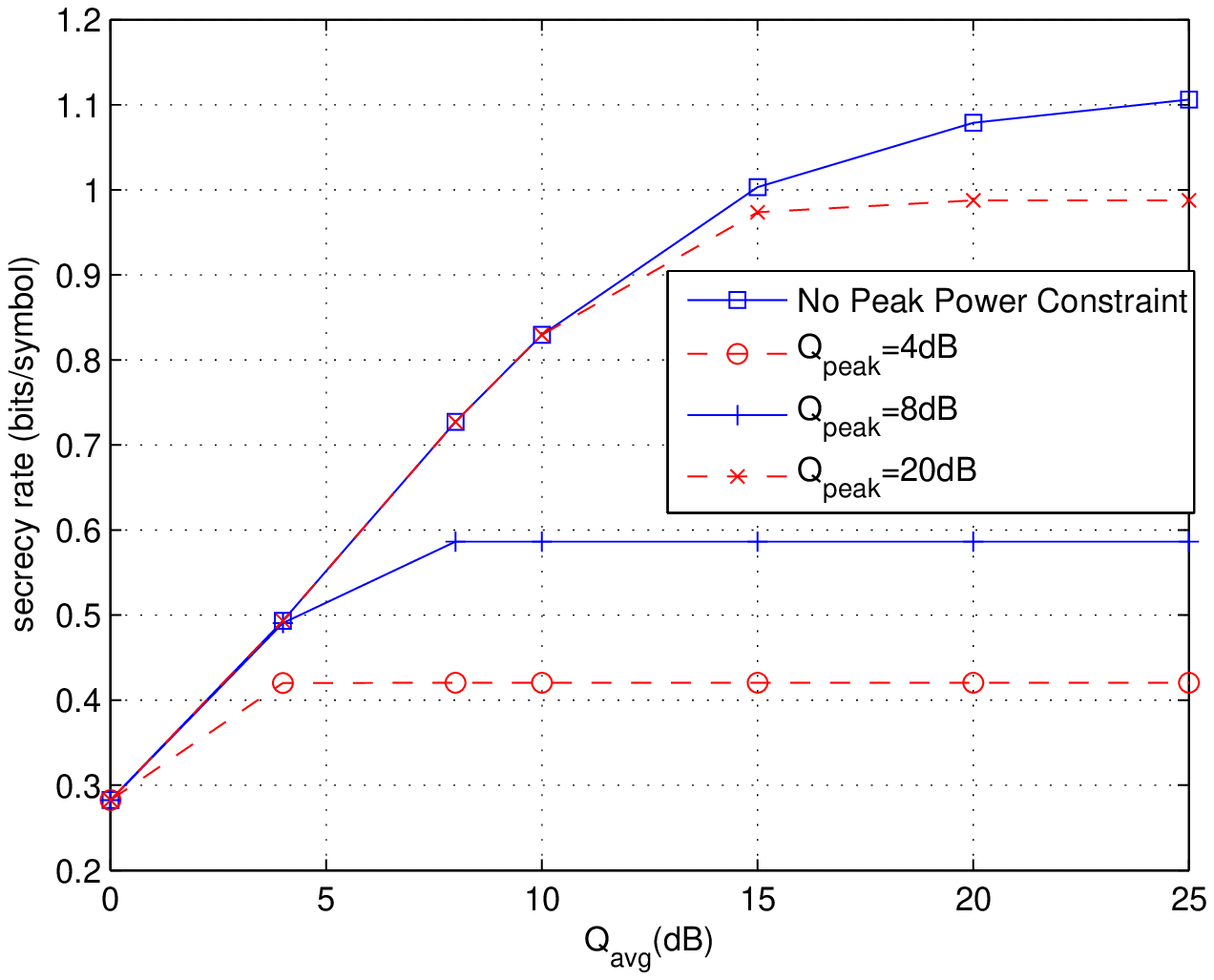}
\caption{secrecy rate vs. $Q_{avg}$ for different peak power
constraint with global CSI available, $\gmbar=\gwbar=1,\gpbar=2$. } \label{fig:2}
\end{center}
\end{figure}

We first consider the case in which the global CSI is available. In Fig.
\ref{fig:2}, we plot the secrecy rate versus $Q_{avg}$ for different values of the peak received
power constraint $Q_{peak}$. We can see from
the figure that, as expected, the larger the $Q_{peak}$, the closer the rate is to
the case of no peak power constraint. We also observe that the
constraint on the peak received power does not have much impact on
the secrecy rate for low values of $Q_{avg}$. On the other hand, as
the value of the average received power limit approaches the peak
received power constraint, the rate plots become flat and the performance gets essentially limited by
the peak received-power constraint.

\begin{figure}
\begin{center}
\includegraphics[width = 0.45\textwidth]{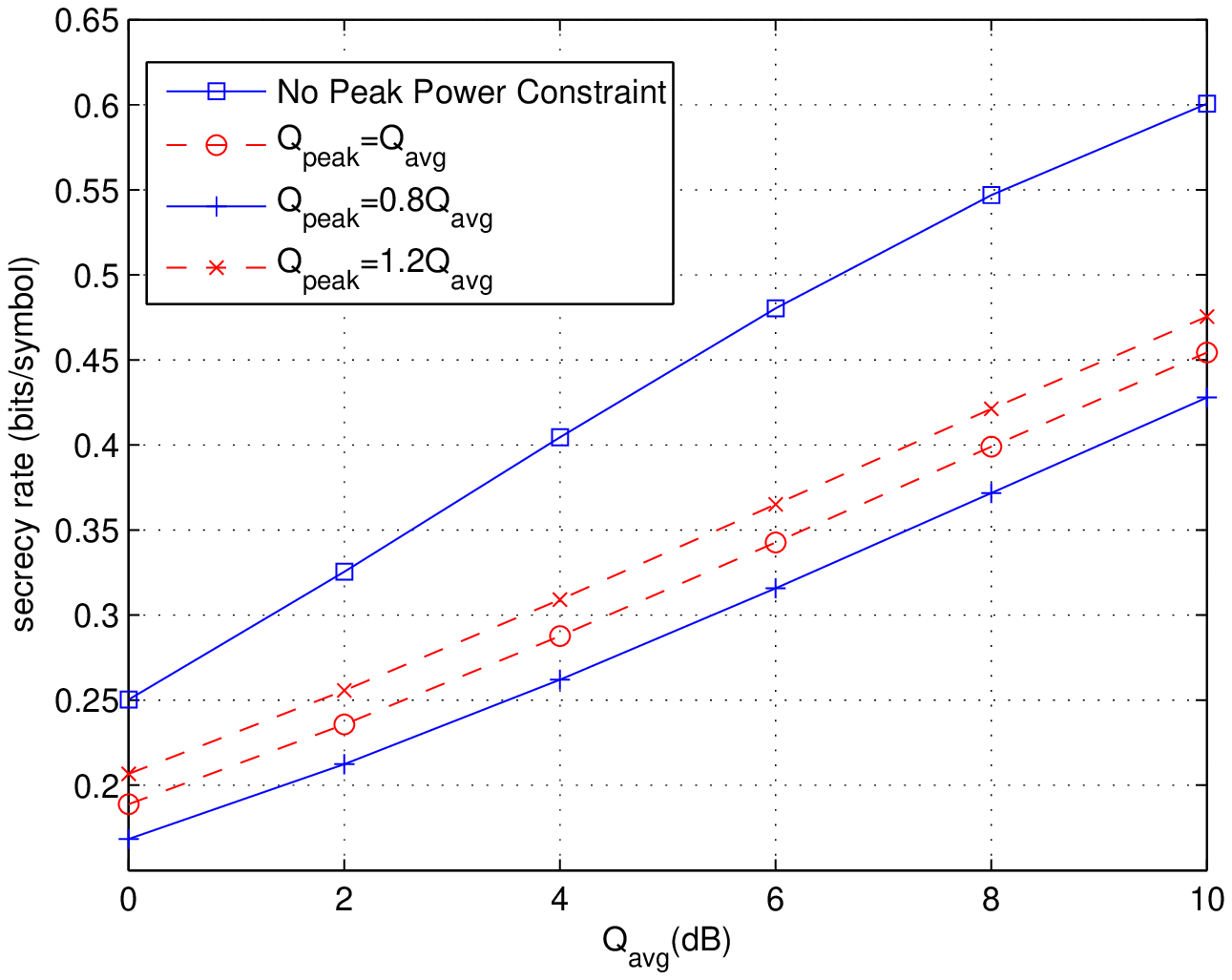}
\caption{secrecy rate vs. $Q_{avg}$ for different peak power
constraint with global CSI available, $\gmbar=1, \gwbar=2,\gpbar=2$.
} \label{fig:3}
\end{center}
\end{figure}

In Fig. \ref{fig:3}, we plot the ergodic secrecy rate as a function
of $Q_{avg}$ while keeping the ratio
$\frac{Q_{peak}}{Q_{avg}}$ fixed. We should point out that
eavesdropper's channel is  stronger than the main channel on average (i.e., $\gmbar=1 < \gwbar=2$) in this figure. Note that
positive secrecy rate can not achieved without fading in such a case. In the figure, we again see
that the higher the ratio $\frac{Q_{peak}}{Q_{avg}}$, the closer the
curve is to the no peak power constraint case. Also, since the peak power constraint becomes more relaxed with increasing $Q_{avg}$, we do not see the flattening of the rate curve in contrast to what is observed in Fig. \ref{fig:2}.

\begin{figure}
\begin{center}
\includegraphics[width = 0.45\textwidth]{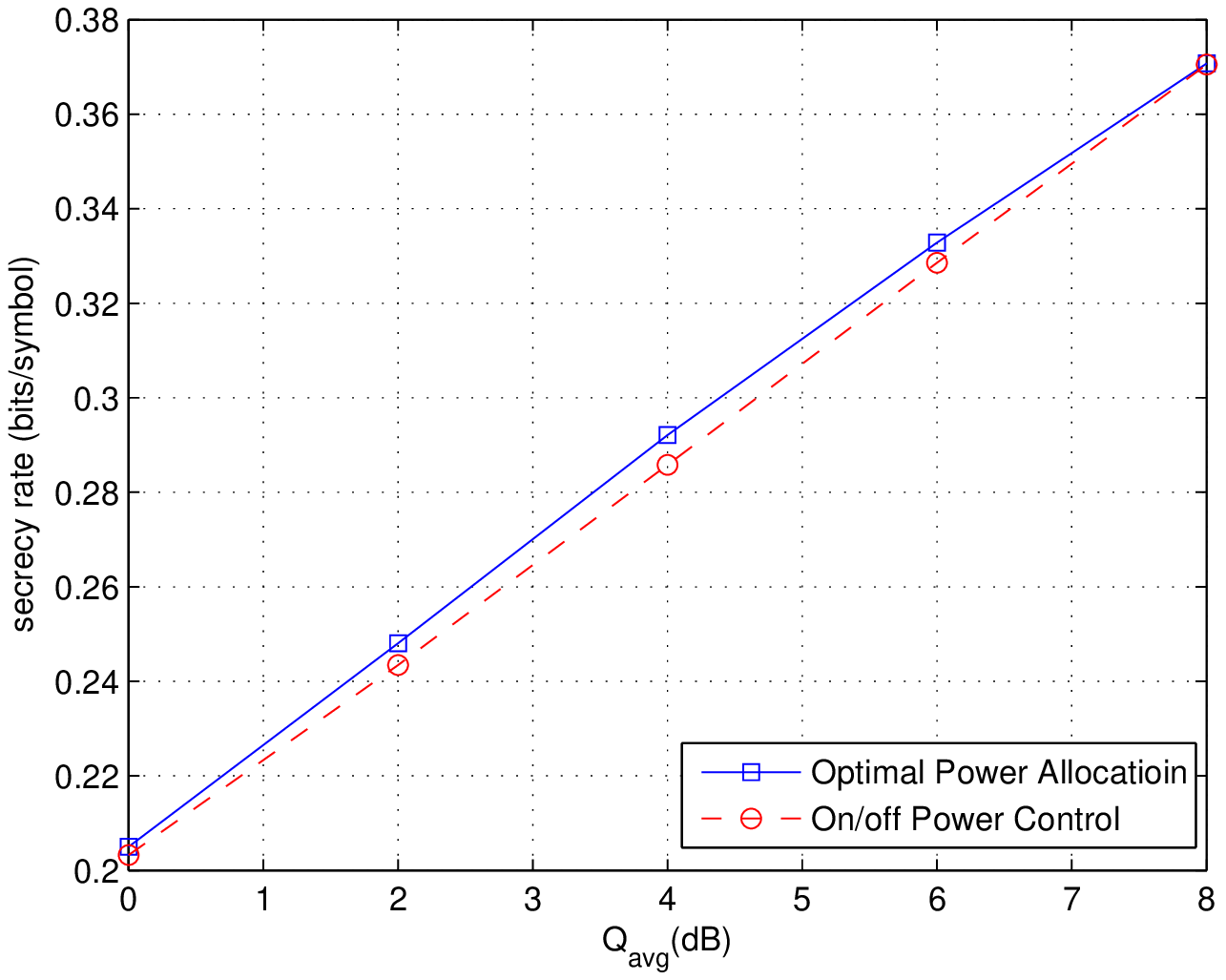}
\caption{secrecy rate vs. $Q_{avg}$ without eavesdropper's CSI,
$\gmbar=1, \gwbar=2,\gpbar=2$. } \label{fig:4}
\end{center}
\end{figure}

Next, we consider the case in which the eavesdropper's CSI is not
available. In Fig.\ref{fig:4}, we plot the ergodic secrecy rate vs. $Q_{avg}$ curves achieved with optimal power allocation and with
the on/off power control method. The fading variances $\bar{\lambda}$ are the same as
in Fig. \ref{fig:3}. By comparing the secrecy rates in Fig. \ref{fig:4} with the secrecy rate in Fig. \ref{fig:3} obtained in the absence of peak constraints, we observe
that not having the eavesdropper's channel information result in a
certain loss in the secrecy rate.  We also see that the
performance of the on/off power control scheme is very close to the
optimal secrecy capacity (when only the main channel and primary channel
CSI is available) for a wide range of SNRs, and approach the optimal
rate when SNR is high. Note that the optimality of the on/off power control
scheme at high SNRs has been
proved in \cite{Gopala} for the secrecy fading channel. Thus, the on/off power control method has
great utility in practical systems due to its advantage of simple implementation.

\section{Conclusion}
In this paper, we have considered a spectrum-sharing system subject to security considerations and studied the optimal power
allocation strategies for the secrecy fading channel under average and peak
received power constraints at the primary user. In particular, we have
considered two scenarios regarding the availability of the CSI. When global CSI is available, we have obtained analytical expressions for the optimal power allocation under average and peak received power constraints. When
only main channel's and primary channel's CSI is
available, we have characterized the optimal power allocation as the
solution to a certain equation. We have also derived the analytical
secrecy rate expression for the simplified on/off power control scheme in
this scenario. Numerical results corroborating our theoretical
analysis have also been provided. Specially, it is shown that the constraint
on the peak received power does not have much impact on the
secrecy rate for low values of $Q_{avg}$ as long as the average
power constraints remain active, and that the performance of
the suboptimal on/off power control scheme approaches the optimal performance
when the eavesdropper's CSI is not available.

\end{document}